\documentclass[reprint,showpacs,superscriptaddress,twocolumn,amsmath,amssymb,aps,pre]{revtex4}

\usepackage{graphicx} 
\usepackage{dcolumn}
\usepackage{bm}

\begin{document}

\title{Bistability and chaos in Taylor-Green dynamo}

\author{Rakesh Yadav}
\affiliation{Department of Physics, Indian Institute of Technology -- Kanpur 208016, India}
\email[Email : ]{yadav.r.k.87@gmail.com}
\author{Mahendra K. Verma}
\affiliation{Department of Physics, Indian Institute of Technology -- Kanpur 208016, India}
\author{Pankaj Wahi}
\affiliation{Department of Mechanical Engineering, Indian Institute of Technology -- Kanpur 208016, India}

\begin{abstract}
Using direct numerical simulations we study dynamo action under the Taylor-Green forcing with Prandtl number less than one.   We observe bistability with a weak magnetic field branch and a strong magnetic field branch.  Both the dynamo branches undergo subcritical dynamo transition.  We also observe host of dynamo states including constant, periodic, quasiperiodic, and chaotic magnetic fields.  One of the chaotic state originates through a quasiperiodic route with phase locking, while another chaotic attractor appears to follow Newhouse-Ruelle-Takens route to chaos.  We also observe intermittent transitions among quasiperiodic and chaotic states for a given Taylor-Green forcing. 
\end{abstract}


\maketitle

\section{Introduction}

Dynamo theory has been applied to explain the generation and  properties of the  magnetic 
fields present in celestial bodies~\cite{Moffat:book, Krause:book, Brandenburg:PR2005}.  In this mechanism, a small magnetic field fluctuation is amplified by the currents induced by the motion of the conducting fluid.   One of the important problems of dynamo research is the nature of ``dynamo transition" from pure fluid state to dynamo state.  The linearized magnetohydrodynamic (MHD) equations yield zero magnetic field in the steady state.  Therefore, the dynamo transition is through a nonlinear instability.   As a result, the nature of transition is very rich exhibiting host of interesting behaviour including subcritical bifurcations~\cite{Ponty:PRL2007, Nigro:ARXIV2010},  variety of dynamo states~\cite{Monchaux:PRL2007, Yadav:EPL2010}, multiple coexisiting attractors~\cite{Yadav:EPL2010} etc.  It has been found that the nature of dynamo onset depends critically on various system properties, e.g., the magnetic Prandtl number (ratio of the kinematic viscosity and the magnetic diffusivity), forcing  function, system geometry, rotation frequency, etc.    Note that the magnetic Prandtl number ($Pm$) of liquid metals, and of the convective fluids of Earth's outer core and the Sun is very small  (of the order of $10^{-5}$), while that of the intergalactic medium is very large (around $10^{14}$).   In the present paper we explore the nature of dynamo transition for a Prandtl number less than unity, with a hope that it could be a representative of low-Prandtl number dynamo.

One of the strong motivations of the dynamo research is to understand the nature of geodynamo and solar dynamo. To this aim, scientists attempt to address the possible dynamo states, and the extent of the induced magnetic field in these systems.  The magnetic Reynolds number $Rm$ (ratio of magnetic advection and magnetic diffusion) of the geodynamo is estimated to be around 125~\cite{Roberts:RMP2000}, which is somewhat near the dynamo transition, so we expect our present work on dynamo transition to be relevant for geodynamo.   The ratio of the magnetic energy and the kinetic energy for the geodynamo is large~\cite{Roberts:RMP2000}, hence it is called strong field dynamo.  Roberts~\cite{Roberts:GAFD1988} argued that rotating sphere with magnetoconvection could show bistability with a weak field branch and a strong field branch.    Kuang and Bloxham~\cite{Kuang:JCP1999} numerically simulated convective dynamo in spherical geometry with rotation and observed weak-field dynamo solution in a simplified system and strong-field dynamo solutions in a more realistic system.    In the present paper we will show that a similar bistability is exhibited in the box geometry without rotation or convection.   We also observe that the transition for both the branches are subcritical.

The magnetic field of the Earth is temporally and spatially chaotic~\cite{Roberts:RMP2000}.   The solar magnetic field is spatially random, but the sun-spot cycle appears to indicate that the primary dipolar field is quasiperiodic~\cite{Ossendrijver:AAR2003}.  In the von K\'arm\'an sodium (VKS) experiment, Monchaux {\it et al.}~\cite{Monchaux:PRL2007} and P\'etr\'elis {\it at al.}~\cite{Petrelis:GAFD2007}  reported various dynamo states including constant, periodic, quasiperiodic, and chaotic magnetic fields.  In the present paper we simulate dynamo transition for a box geometry, and report various dynamo states (including chaos) similar to those obtained in the VKS experiment.   

We focus on the behaviour of Taylor-Green (TG) dynamo for $Pm=0.5$ fluid in order to probe dynamo behaviour for low magnetic Prandtl number (low-$Pm$) regime. Numerical simulation of very low-$Pm$  fluid (around $10^{-5}$ corresponding to liquid metals) is very difficult since the corresponding Reynolds number for dynamo transition for such fluid is more than $10^6$~\cite{Monchaux:PRL2007}; simulation of such high Reynolds number would require grid resolution far more than presently achievable in the modern supercomputers.  We adopt TG forcing since it has somewhat similar flow structure as the VKS experiment.   

Earlier, Nore {\it et al.}~\cite{Nore:PP1997} simulated TG dynamo for $Pm$ near unity.  Ponty {\it et al.}~\cite{Ponty:PRL2005} studied dynamo mechanism for low magnetic Prandtl numbers ($Pm$ $\approx$ 1 to $10^{-2}$) using hyperdissipative parameters and large-eddy simulations, and observed a sharp increase in the critical magnetic Reynolds number (threshold of dynamo transition) with the decrease of $Pm$.  Ponty {\it et al.}~\cite{Ponty:PRL2007}  observed a subcritical dynamo bifurcation for the TG dynamo in the low-$Pm$ regime.  Mininni {\it et al.}~\cite{Mininni:APJ2005} and  Yadav {\it et al.}~\cite{Yadav:EPL2010} studied the energy transfers, and the geometry of the velocity and the magnetic field structures. Yadav {\it et al.}~\cite{Yadav:EPL2010} also observed a supercritical pitchfork bifurcation for the dynamo transition for $Pm=1$.  They reported a large number of dynamo states including constant, periodic, quasiperiodic, and chaotic magnetic fields. Dubrulle {\it et al.}~\cite{Dubrulle:NJP2007} investigated various bifurcations in both hydrodynamic (with no magnetic field)  and magnetohydrodynamic systems under TG forcing.   Scientists have also studied dynamo behaviour for various kinds of forcing, e.g.,  Roberts, ABC, Ponomarenko,  and random, as well as for different geometries, e.g., box, cylinder, sphere etc.~\cite{Dormy:book}. Dynamo transition and subsequent dynamo states have also been studied using low-dimensional models of dynamo~\cite{Weiss:GAFD2010}.

The outline of the paper is as follows:  The numerical procedure is described briefly in Section \ref{sim_meth}.  Bifurcation analysis is presented in Section~\ref{results}.  We analyze several chaotic windows and the routes to chaos in Section~\ref{routes}. We conclude in Section~\ref{conclusions}.

\section{\label{sim_meth}Simulation Methodology} 
The governing equations of dynamo are same as those of MHD, which are
\begin{eqnarray}
\partial_{t}\mathbf{u}+ (\mathbf{u} \cdot \nabla) \mathbf{u} & = &
-\nabla p+ (\mathbf{J} \times \mathbf{B} ) + \nu\nabla^{2}\mathbf{u}+\mathbf{F}, \label{eq:MHD_vel}\\
\partial_{t}\mathbf{B} & = &
\nabla \times ( \mathbf{u} \times \mathbf {B} )+\eta\nabla^{2}\mathbf{B}, \label{eq:MHD_mag}\\
\nabla \cdot \mathbf{u} & = & 0,  \label{eq:div_v_0} \\
\nabla \cdot \mathbf{B} & = & 0,  \label{eq:div_B_0}
\end{eqnarray}
where $\mathbf{u}$ is the fluid velocity, $\mathbf{J}$ is the
current density, $\mathbf{B}$ is the magnetic field, $p$ is the 
hydrodynamic pressure, $\nu$ is the kinematic viscosity, $\eta$ is the 
magnetic diffusivity, and $\mathbf{F}$ is the external force field. The density of the fluid is assumed to be unity. The three important parameters related to dynamo instability are the magnetic Prandtl number $Pm = \nu/\eta$, the Reynolds number $Re = UL/\nu$, and the magnetic Reynolds number $Rm = UL/\eta$, where $U$ and $L$ are the large velocity scale and the large length scale, respectively.  Note that $Rm = Re \times Pm$, hence only two of the above three parameters are independent.    In our study, we fixed the magnetic Prandtl number to 0.5 with $\nu = 0.1$ and $\eta = 0.2$.

We solve the MHD equations (Eqs.~(\ref{eq:MHD_vel}-\ref{eq:div_B_0})) numerically for a box geometry of size $(2\pi)^3$ with  periodic boundary conditions in all the directions. We use a pseudospectral code TARANG~\cite{TARANG} to carry out our simulations.   We apply fourth-order Runge-Kutta  scheme for time advancement with dynamically adjusted $dt$ chosen using the CFL condition: $dt = \Delta x/\sqrt{20 E^u}$ where $\Delta x$ is the grid spacing and $E^u$ is the total kinetic energy.   On the velocity field we apply Taylor Green forcing 
\begin{eqnarray}
\mathbf{F}(k_0) & = & F_0
\left[
\begin{array}{c}
\sin(k_0 x)\cos(k_0 y)\cos(k_0 z) \\
-\cos(k_0 x)\sin(k_0 y)\cos(k_0 z) \\
0
\end{array}
\right], \label{eq:taylor_green}
\end{eqnarray}
where $F_0$ is the amplitude of the forcing and $k_0$ is the wavenumber. We set $k_0$ equal to 2. Note that the TG forcing has components only along $x$ and $y$ directions.  Numerical 
simulations reveal that the TG forcing induces counter rotating eddies, and it  mimics the flow structure of the VKS experiment qualitatively~\cite{Monchaux:PRL2007}. 

The  box is discretized uniformly in all the directions, with most of the simulations on $64^3$ grid. This resolution was sufficient to resolve our simulation near the dynamo transition, as demonstrated by the fact that the product of the Kolmogorov length and the largest wavenumber lies between $1.3 - 9$ for all our runs.  We also verified the grid-independence by performing few runs of a given dynamo state on $128^3$ and $64^3$ grids.   All our simulation were dealiased using the 2/3 rule.  We performed approximately 150 simulations for various forcing parameters ($F_0 = 1:46$), and studied the global kinetic and magnetic energies, as well as the 
amplitudes of the velocity and magnetic Fourier modes. The magnetic Reynolds number in our 
simulations ranged from around $3$ to $90$.

The importance of large scale modes have been amply highlighted in dynamo literature.  In our present dynamo simulation, as well as for $Pm=1$ reported earlier by Yadav {\it et al.}~\cite{Yadav:EPL2010},  some of the magnetic and kinetic Fourier modes play a dominant role.  For the magnetic Reynolds numbers employed in these simulations, the  dominant velocity  modes  are  $(\pm 2, \pm 2, \pm 2)$, 
$(\pm 4, \pm 4, \pm 4)$,  $(\pm 4, \pm 4, 0)$, and the dominant magnetic modes are 
$(0, 0, \pm 1$), $(0, 0, \pm 2)$, $(0, 0, \pm 3)$, $(\pm 2, \pm 2, \mp 3)$, 
$(\mp 2, \mp 2, \pm 1)$.
Here, the three arguments refer to $x$, $y$, and $z$ components of a wavenumber in the Fourier space.   The  $(\pm 2, \pm 2, \pm 2)$ velocity mode contains maximum kinetic energy due to the TG forcing at  $k=(2,2,2)$.   The above set of modes carry more than 95\%  of the total energy.   Note, however, that higher wavenumber velocity and magnetic Fourier modes become important in the turbulent dynamo.  We also point out that the above modes play similar roles as the large-scale modes used in the earlier dynamo literature~\cite{Petrelis:JPCM2008, Moffat:book}.

In the present paper we analyze the  time series and phase space plots of the Fourier modes.  This exercise provides us with information about various dynamo states as well as the nature of bifurcations.    Primary and secondary instabilities can be conveniently illustrated using bifurcation diagrams.  In the next section we will describe bifurcation diagrams for $Pm = 0.5$ constructed using some of the dominant magnetic Fourier modes.

\section{\label{results}Bifurcation analysis \& strong and weak dynamo branches}

\begin{figure}
\begin{center}
\includegraphics[scale=0.58]{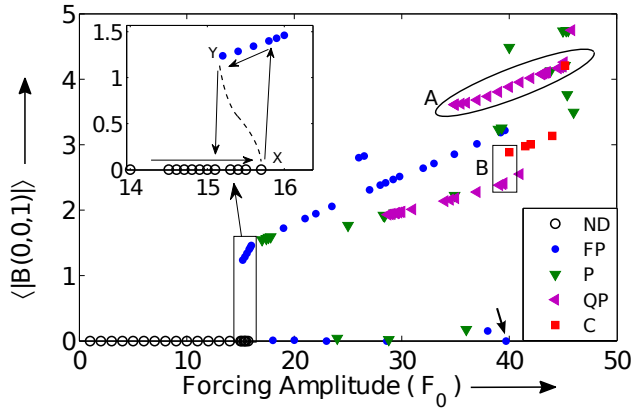}
\caption{Bifurcation diagram: Time averaged values of ${|{\bf B}(0,0,1)|}$  for various forcing amplitude $F_0$.  The figure illustrates various dynamo states with ND = no-dynamo state, FP = fixed point (constant in time), P = periodic state, QP = quasiperiodic state, and C = chaotic state. The inset for $F_0 = 14$:$16$ illustrates a sudden jump and a hysteresis loop indicating that the transition is subcritical. Also, oval `A' exhibits  quasiperiodic route to chaos (see Subsection~\ref{PL}), and the rectangular box `B' exhibits Newhouse-Ruelle-Takens route to chaos (see Subsection~\ref{NRT}).}
\label{fig:states}
\end{center}
\end{figure}

We construct a bifurcation diagram by using the time averaged values of the magnitude 
of the magnetic Fourier mode ${\bf B}(0,0,1)$ for different values of $F_0$ (see Fig.~\ref{fig:states}).  These values are computed after the system reaches a steady state.  Since the Fourier amplitudes are generally complex, it is convenient to use the absolute value of a Fourier mode to depict the dynamo states. 

The inset of Fig.~\ref{fig:states} shows a zoomed view of the dynamo transition region near $F_0 = 15.8$. We observe fluid or no-dynamo state for $F_0 < 15.8$.  At $F_0 = 15.8$, we obtain a dynamo state with $|{\bf B}(0,0,1)|$  as well as the total magnetic energy ($E^b$) showing a finite jump.  However, when we use the dynamo state at $F_0=15.8$ as our initial condition and gradually decrease the forcing amplitude, as shown in the inset of 
Fig.~\ref{fig:states}, the dynamo state continues till $F_0\approx15.2$, at which 
point there is a sudden jump to the fluid state.  This feature of hysteresis 
for $F_0 = 15.2:15.8$ demonstrates the subcritical nature of the transition, with a  ``subcritical pitchfork bifurcation" at X and a ``saddle-node bifurcation" at Y (see the inset of the figure). These points are joined together using a dashed curve to depict the unstable branch.  In a recent study, Krstulovic {\em et al.}~\cite{Krstulovic:ARXIV2011} also observed subcritical dynamo transition in their simulation and low-dimensional model.   In a planetary dynamo context, Kuang {\em et al.}~\cite{Kuang:GRL2008} argued that the sudden termination of Martian dynamo could be due to subcriticality. 

\begin{figure}
\begin{center}
\includegraphics[scale=0.50]{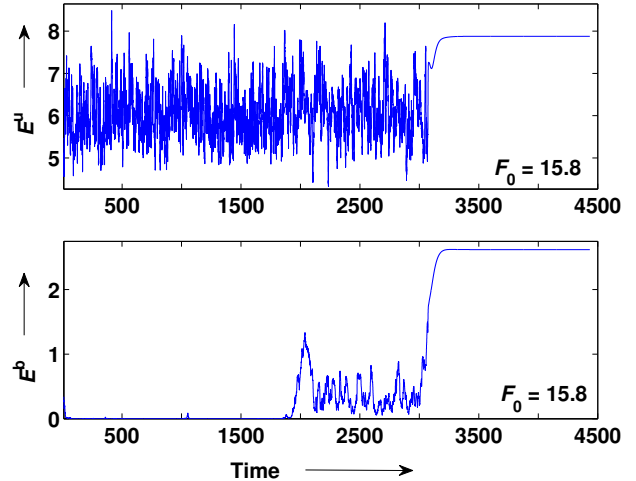}
\caption{The total kinetic energy (top panel) and the total magnetic 
energy (bottom panel) for a forcing amplitude of 15.8 indicating a constant value for the steady state.}
\label{fig:TS_KE_ME_15.8}
\end{center}
\end{figure}

Another noteworthy feature during the transition at $F_0=15.8$ is that the Reynolds number is sufficiently high ($\approx 60$) for the fluid  to be temporally chaotic.  However, in the corresponding dynamo states at the same $F_0$, both the velocity and magnetic fields become constant in time, as shown in Fig.~\ref{fig:TS_KE_ME_15.8}.  This is due to the presence of the newly-born finite magnetic field. Morin {\it et al.}~\cite{Morin:IJMPB2009} reported similar effects of magnetic field for spherical shell dynamo. After $F_0 \approx 16.2$, the dynamo solution bifurcates to  periodic states, and subsequently to quasiperiodic and chaotic states as shown in Fig.~\ref{fig:states}.

\begin{figure}
\begin{center}
\includegraphics[scale=0.58]{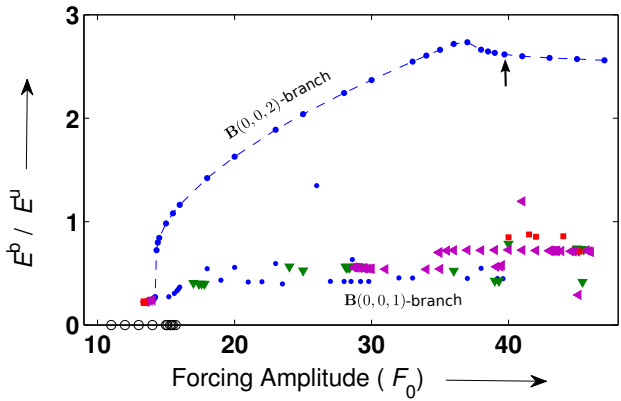}
\caption{Bifurcation diagram: Ratio of the total magnetic energy and the total kinetic energy ($E^b$/$E^u$) $\it vs.$  the forcing amplitude ($F_0$). Description of different symbols is provided in Fig.~\ref{fig:states}. The ${\bf B}(0,0,2)$-branch was constructed using the dynamo state marked with an arrow as initial condition.}
\label{fig:Eb_by_Eu_vs_F}
\end{center}
\end{figure}

\begin{figure}
\begin{center}
\includegraphics[scale=0.58]{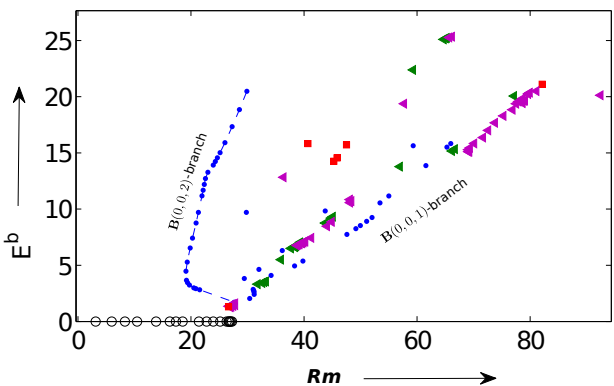}
\caption{Bifurcation diagram: Total magnetic energy ($E^b$) $\it vs.$ the magnetic Reynolds number ($Rm$). Description of different symbols is provided in Fig.~\ref{fig:states}. }
\label{fig:ME_vs_Rm}
\end{center}
\end{figure}

For some of our dynamo runs, $\langle {|{\bf B}(0,0,1)|}$ is negligible, as shown by the dots on the $x$ axis of Fig.~\ref{fig:states}.   For these simulations, the magnetic mode ${\bf B}(0,0,2)$ becomes dominant;  we denote the set of these states as the ``${\bf B}(0,0,2)$-branch". The other collection of states with ${\bf B}(0,0,1)$ as the most dominant mode is referred to as the ``${\bf B}(0,0,1)$-branch".  We observe that the properties of these two branches are quite different.   
We illustrate this feature using another  bifurcation diagram whose vertical axis is  $E^b/E^u$, the ratio of the magnetic energy and the kinetic energy (shown in Fig.~\ref{fig:Eb_by_Eu_vs_F}).   This bifurcation diagram reveals two distinct branches:  the upper curve is the ${\bf B}(0,0,2)$-branch, for which the magnetic field remains constant in time, and the lower one is the ${\bf B}(0,0,1)$-branch.  The ${\bf B}(0,0,2)$ branch was constructed using the fixed point solution at $F_0=39.7$, which is marked with small arrows in Figs.~\ref{fig:Eb_by_Eu_vs_F} and ~\ref{fig:states}, as initial condition.  Also, the ratio $E^b/E^u$ for the ${\bf B}(0,0,2)$-branch is larger than the corresponding ratio for the ${\bf B}(0,0,1)$-branch.  Thus we address the former as a ``strong-field branch", and the latter as a ``weak-field branch".   The above feature regarding the strength of the magnetic field is corroborated in  Fig.~\ref{fig:ME_vs_Rm} where we plot the magnetic energy ($E^b$) vs. the magnetic Reynolds number ($Rm$).    The two distinct branches described above demonstrate that the TG dynamo at $Pm=0.5$ exhibits ``bistability", similar to those reported by Simitev and Busse~\cite{Simitev:EPL2009}.

\begin{figure}
\begin{center}
\includegraphics[scale=0.53]{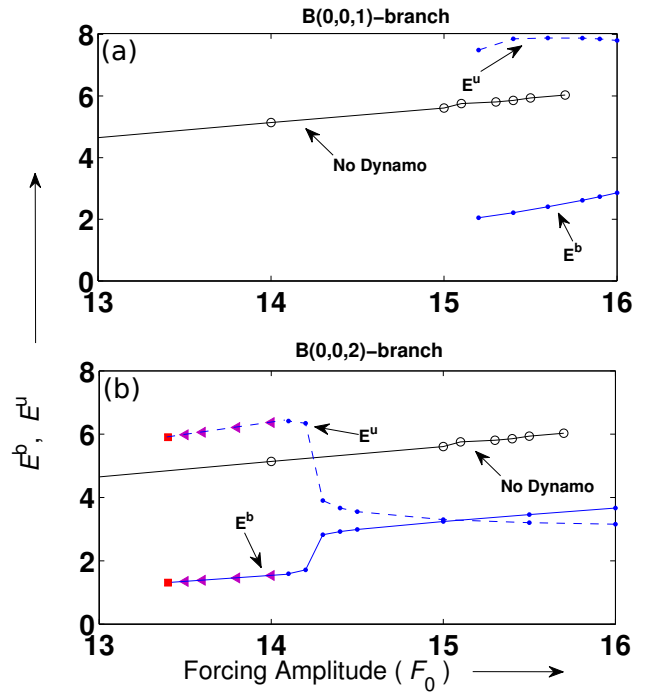}
\caption{A plot of the total kinetic energy ($E^u$) and the total magnetic energy ($E^b$) $\it vs.$ 
the $F_0$ near the dynamo transition. Description of different symbols is provided in Fig.~\ref{fig:states}. (a): The ${\bf B}(0,0,1)$-branch portrays the subcritical nature of the transition.  (b):  The ${\bf B}(0,0,2)$-branch illustrates transition from a fixed point state to quasiperiodic (purple triangles) and subsequently to chaotic state (red square) as $F_0$ is decreased.  }
\label{fig:Eu_Eb_vs_F}
\end{center}
\end{figure}

The dynamo transition for the ${\bf B}(0,0,2)$-branch differs significantly from that of the ${\bf B}(0,0,1)$-branch described earlier.   For the ${\bf B}(0,0,1)$-branch, the jumps in the kinetic and magnetic energy shown in  Fig.~\ref{fig:Eu_Eb_vs_F}(a) is consistent with the subcritical nature of the transition described earlier.  However, the ${\bf B}(0,0,2)$-branch exhibits chaos at the dynamo transition itself. The origin of the chaotic state becomes apparent when we decrease $F_0$ from the fixed point solution.   As shown in Fig.~\ref{fig:Eu_Eb_vs_F}(b), the fixed point bifurcates to quasiperiodic (purple triangles of the figure) and subsequently to chaotic state (red squares) as the $F_0$ is decreased.   The route to chaos for the ${\bf B}(0,0,2)$-branch may be similar to those observed by Pal {\it et al.}~\cite{Pal:EPL2009} for zero-Prandtl number convection.   The quasiperiodic and the chaotic states are clustered in Fig.~\ref{fig:Eb_by_Eu_vs_F} near the transition.

Earlier, the weak field branch and strong field branch of dynamo have been reported in the context of convection-driven hydromagnetic dynamos~\cite{Childress:PRL1972, Roberts:GAFD1988, Kuang:JCP1999,Morin:ARXIV2011}. Pierre~\cite{Pierre:collec} and Stellmach and Hansen~\cite{Stellmach:PRE2004} numerically demonstrated this phenomenon in Childress-Soward dynamo where they reported that the Strong-field branch goes even further back than the weak-field branch in the parameter space, analogous to the ${\bf B}(0,0,2)$-branch reported in this paper. Similar behaviour has also been reported by Sreenivasan and Jones~\cite{Sreenivasan:JFM2011} in a rapidly rotating spherical
shell dynamo.  It is interesting that our system in the box geometry, without any convection or rotation, exhibits weak and strong dynamo branches.  

In the next section we will focus on several chaotic dynamo states and study routes to chaos for these states.

\section{\label{routes}Routes to Chaos}

The phase space portraits drawn using the Fourier modes reveal multiple windows of chaos near the dynamo transition itself.    In the present paper we describe routes to chaos for two chaotic windows: the oval-shaped enclosure `A' and  the rectangular-shaped box `B'  shown in Fig.~\ref{fig:states}.  In the following discussion we show that the origin of chaos for the dynamo states of box `A' follow a quasiperiodic route to chaos through phase locking, while the corresponding route to chaos for the box `B'  is through  the Newhouse-Ruelle-Takens scenario.  

We study the routes to chaos using  some of the dominant modes of magnetic field, viz. ${\bf B}(0,0,1)$, ${\bf B}(0,0,2)$, and ${\bf B}(0,0,3)$. We use time series, phase-space projections, and 
Poincar\'e sections for this study~\cite{Hilborn:book}. 

\subsection{\label{PL}Quasiperiodic route to chaos through phase-locking}

\begin{figure}
\begin{center}
\includegraphics[scale=0.59]{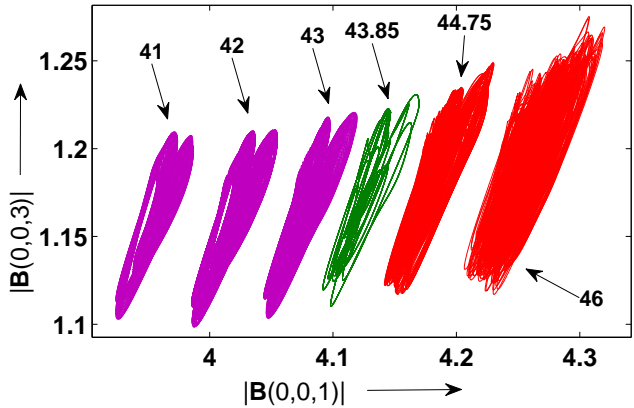}
\caption{Phase space projections on the $|{\bf B}(0,0,1)|$-$|{\bf B}(0,0,3)|$
plane demonstrating a quasiperiodic  route to chaos through phase-locking. The forcing amplitudes ($F_0$) for different attractors are marked with arrows.   $F_0$ = 41, 42, 43 (pink), $F_0$ = 43.85 (green), and $F_0$ = 44.75, 46 (red) correspond to quasiperiodic, phase-locked, and chaotic states, respectively.}
\label{fig:phase_locking}
\end{center}
\end{figure}

\begin{figure}
\begin{center}
\includegraphics[scale=0.49]{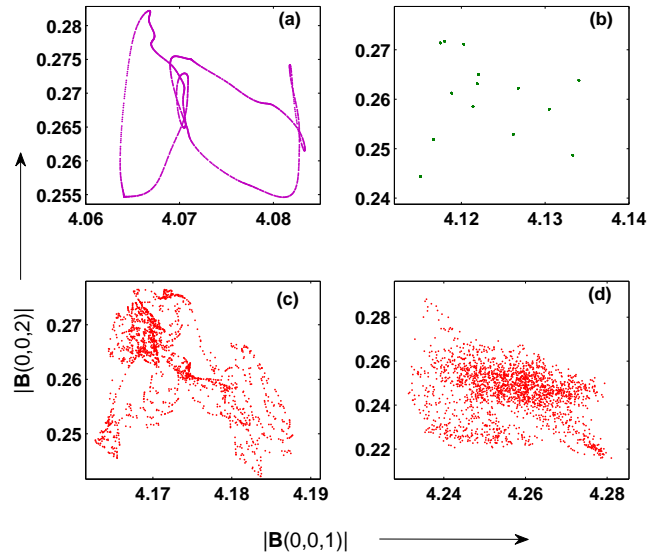}
\caption{Poincar\'e sections for some of the attractors of Fig.~\ref{fig:phase_locking} with $|{\bf B}(0,0,3)|$ = 1.15 as the Poincar\'e plane.  (a): quasiperiodic attractor for $F_0 = 43$, (b):  phase-locked attractor for $F_0= 43.85$, (c,d): chaotic attractors for $F_0 = 44.75$ and  46.} 
\label{fig:PS_PL}
\end{center}
\end{figure}

In Fig.~\ref{fig:phase_locking} we illustrate the phase space projections of dynamo states on 
$|{\mathbf B}(0,0,1)|$-$|{\mathbf B}(0,0,3)|$ plane for the forcing range of $F_0 = 41$:46 
(corresponding to the oval-shaped enclosure `A' of Fig.~\ref{fig:states}).  For $F_0=41,42$, and 43, the system is  quasiperiodic since the phase space projection is densely filled up. The approximate values of the two incommensurate frequencies for the $F_0=43$ dynamo state are 0.0165 and 0.0208 (for $|{\mathbf B}(0,0,1)|$ time series). For $F_0=43.85$, the system becomes periodic or phase-locked.  The time period 
of the periodic orbit is relatively large.   The emergence of periodic orbit after 
quasiperiodic solutions is called ``phase locking"~\cite{Hilborn:book}. A subsequent  
increase of $F_0$ leads to a chaotic state, as evident from the phase space projections for 
$F_0=44.75$ and 46.  The nature of attractors is corroborated by the Poincar\'{e} 
sections for $F_0=43, 43.85, 44.75$, and 46 presented in Fig.~\ref{fig:PS_PL}
(a, b, c, d), respectively.  These Poincar\'e sections were obtained by using the $|{\bf B}
(0,0,3)|=1.15$   as the Poincar\'e intersection plane for the phase space trajectories 
in the subspace of $|{\bf B}(0,0,1)|$-$|{\bf B}(0,0,2)|$-$|{\bf B}(0,0,3)|$.   Thus the 
route to chaos for the chaotic trajectories in the oval-box `A' of Fig.~\ref{fig:states} 
is through the quasiperiodic route via phase locking, first reported for 
circle map~\cite{Hilborn:book}.

\subsection{\label{NRT}Newhouse-Ruelle-Takens scenario}

We observe an interesting set of dynamo states in the rectangular-box `B' of Fig.~\ref{fig:states}.  Fig.~\ref{fig:Phase_space_2F_3F_chaos}(a) illustrates the phase space projections on the $|{\bf B}(0,0,1)|$-$|{\bf B}(0,0,3)|$ plane for $F_0 = 39.3$ and 39.464, corresponding to the two purple triangles in the rectangular-box `B' of Fig.~\ref{fig:states}.  Also, the power spectral density plot (PSD) of the $|{\bf B}(0,0,1)|$ time series of these two dynamo states is shown in Fig.~\ref{fig:combined_PSD}(a,b). The PSD reveals that the state at $F_0 = 39.3$ contains two incommensurate frequencies $f_1$ and $f_2$ (Fig.~\ref{fig:combined_PSD}(a)), while the state at $F_0 = 39.464$ has three incommensurate frequencies $f_1$, $f_2$, and $f_3$ (Fig.~\ref{fig:combined_PSD}(b)).  Thus the corresponding dynamo states reside on ``2-torus'' ($T^2$) and ``3-torus" ($T^3$) respectively in the subspace.   We expect that a further increase of $F_0$ should push the system to chaos following a Newhouse-Ruelle-Takens scenario.  However, we have not yet found the corresponding chaotic attractor.  Recently, Stefani {\it et al.}~\cite{Stefani:AN2011} have observed similar quasiperiodic routes to chaos in a dynamo model.  

Instead of transition from $T^3$ to a chaotic dynamo state, a very small increase of $F_0$ pushes the system to a new chaotic attractor, whose span is much larger than those of the $T^2$ or $T^3$.   In Fig.~\ref{fig:Phase_space_2F_3F_chaos}(b), we illustrate the larger chaotic attractor obtained for $F_0=39.51$, corresponding to the red square in the rectangular-box `B' of Fig.~\ref{fig:states}. In Fig.~\ref{fig:Phase_space_2F_3F_chaos}(b),  $T^2$ or $T^3$ states reside in the circled region. The large attractor for $F_0=39.51$ has a different origin as the PSD shown in Fig.~\ref{fig:combined_PSD}(c) is quite different from those in Fig.~\ref{fig:combined_PSD} (a, b).   The origin of the larger attractor is not well understood, however, it is possibly through a ``crisis''.    Note that a large number of attractors exists for this range of $F_0$, as evident from the Fig.~\ref{fig:states}.  Chaos can emerge due to the intersections of multiple attractors or their basins of attraction~\cite{Hilborn:book}.   A detailed investigation of these issues is beyond the scope of this paper. 

We performed dynamo simulations near $F_0=39.51$ and observe interesting features involving intermittent transitions between three attractors; this dynamics will be described in the next subsection.

\subsection{Intermittent transitions between various attractors}

\begin{figure}
\begin{center}
\includegraphics[scale=0.47]{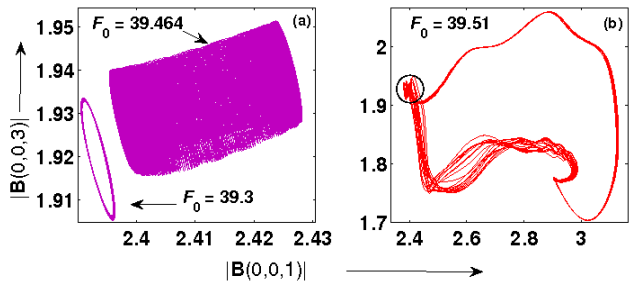}
\caption{Phase space projections on the ($|{\bf B}(0,0,1)|$-$|{\bf B}(0,0,3)|$) 
plane. Panel (a) contains a stable 2-torus and a 
3-torus quasiperiodic state and panel (b) shows a close lying (in parameter space) 
large chaotic attractor. The corresponding forcing amplitude 
is marked with arrow. The circle 
drawn in (b) portrays a rough phase space span of the attractors shown in (a).}
\label{fig:Phase_space_2F_3F_chaos}
\end{center}
\end{figure}

\begin{figure}
\begin{center}
\includegraphics[scale=0.41]{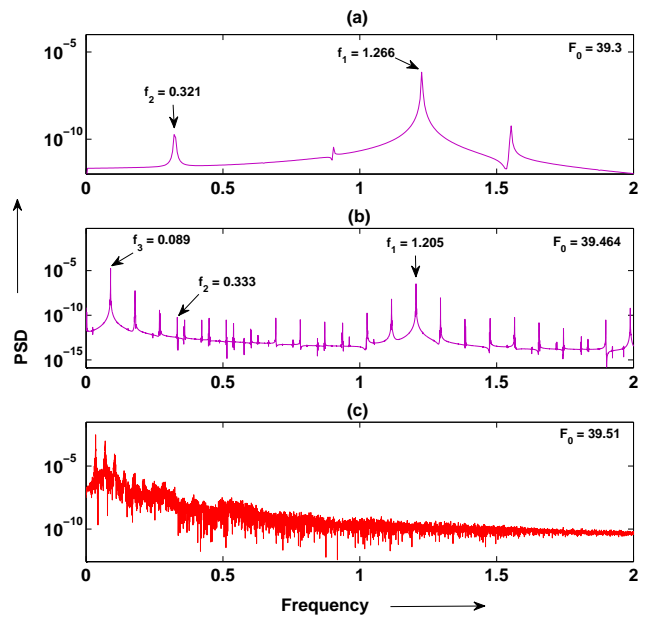}
\caption{Power Spectral Density (PSD) plot of the dynamo states shown in 
Fig.~\ref{fig:Phase_space_2F_3F_chaos}. The dominant frequencies are depicted in  the figures with arrows.  (a): For $F_0=39.3$, the state has two incommensurate dominant frequencies; (b): For $F_0=39.464$, the state has three incommensurate frequencies; (c):  A dense 
power spectrum corresponding to the large chaotic attractor portrayed in Fig.~\ref{fig:Phase_space_2F_3F_chaos}(b).}
\label{fig:combined_PSD}
\end{center}
\end{figure}

\begin{figure*}
\begin{center}
\includegraphics[scale=0.5]{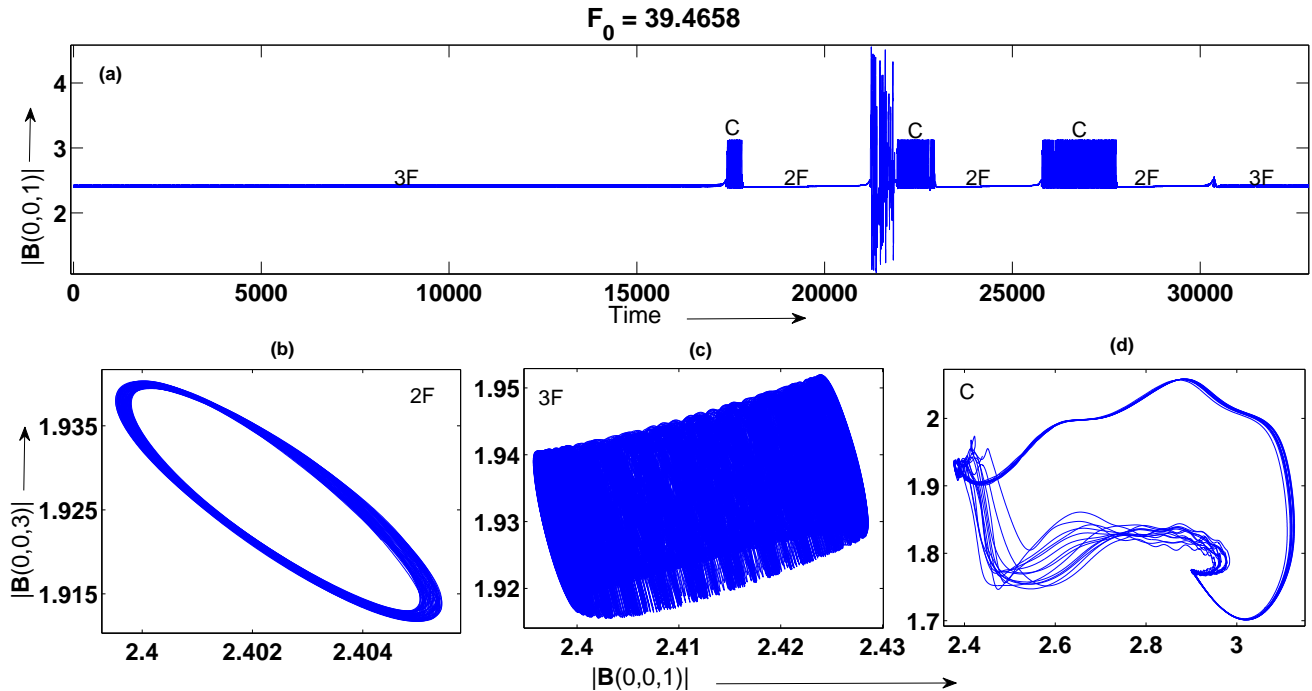}
\caption{Time evolution of a dynamo state at $F_0 = 39.4658$ showing intermittent transitions among three attractors: quasiperiodic state $T^2$(b),  quasiperiodic state $T^3$(c), and chaotic state(d).  The subfigures (b,c,d) are the phase space projections at three different time intervals during the evolution.}
\label{fig:TS_Phase_space_39.4658}
\end{center}
\end{figure*}

For the forcing amplitude $F_0 = 39.4658$ we observe intermittency.  A long time series for $F_0 = 39.4658$ simulation is illustrated in Fig.~\ref{fig:TS_Phase_space_39.4658}(a), which illustrates that the system makes intermittent transitions among three attractors, `2F', `3F', `C', shown below the time series.  Note that the 2F, 3F, and C attractors are qualitatively similar to the state shown in Fig.~\ref{fig:Phase_space_2F_3F_chaos}  obtained for $F_0=39.3, 39.464$, and 39.51, respectively.  Thus, at $F_0 = 39.4658$ the system appears to hop over various attractors.    The intermittent transition among various attractors described above is  similar to ``intermittency" in which a system switches between an ``ordered" state and a chaotic state for a single parameter.   In Fig.~\ref{fig:TS_Phase_space_39.4658}  the quasiperiodic states are the ordered state.   The system has been evolved for $50000$ eddy turnover times, and only a part of this time series has been shown in Fig.~\ref{fig:TS_Phase_space_39.4658}(a).  Hence the observed phenomena is not a transient. The large fluctuations for time near $21250$ in Fig.~\ref{fig:TS_Phase_space_39.4658}(a) are transient fluctuations that settle down quickly.

\section{\label{conclusions}Conclusions and Speculations} 
Our Taylor-Green dynamo simulations for $Pm = 0.5$ near the dynamo transition reveal interesting properties.  We observe bistability, with a  weak magnetic field branch and a strong magnetic field  branch.   The dominant Fourier modes for the two branches are ${\bf B}(0,0,1)$ and ${\bf B}(0,0,2)$ respectively.   Both of these branches have subcritical origin.   Qualitatively, these branches resemble the weak- and strong-field dynamo actions in rotating magnetoconvection reported earlier~~\cite{Roberts:GAFD1988,Kuang:JCP1999}.   Note that the subcritical dynamo transition for $Pm=0.5$ is in sharp contrast with the TG dynamo for $Pm=1$ which shows a supercritical pitchfork bifurcation at the transition~\cite{Yadav:EPL2010}. This observation is consistent with the recent work by Krstulovic {\it et al.}~\cite{Krstulovic:ARXIV2011}.

Our simulations also reveal various kinds of dynamo states including constant, periodic, quasiperiodic, and chaotic magnetic fields.  We analyzed two chaotic windows among several ones observed in our simulations.  Chaos for these windows arise through quasiperiodic route.  One of them undergoes a phase-locking, while the other one appears to be through the  Newhouse-Ruelle-Takens route. We also observe an intermittent transitions between quasiperiodic ($T^2, T^3$) and chaotic attractors.  Various dynamo states reported in our simulations are quite similar to those observed in the VKS experiment~\cite{Monchaux:PRL2007}.

We performed our simulations on a periodic box, which is an idealized geometry.  Yet, we observe similarities with the VKS experiment and geodynamo simulations.  This could be due to similarities in the inherent nonlinearities in the system (independent of the geometry etc.).  Also, the  behaviour of dynamo transition and dynamo states for $Pm=0.5$ discussed in the present paper differs significantly from that for $Pm=1$ studied by Yadav {\em et al.}~\cite{Yadav:EPL2010}.   Extensions of the present work to more realistic geometry and lower magnetic Prandtl numbers would yield interesting insights into this challenging field.

\section{Acknowledgments}
We thank Binod Sreenivasan, Stephan Fauve, and Mani Chandra for fruitful discussions and comments. This work was supported by a research grant of DST India as 
Swarnajayanti fellowship to MKV.   Part of the simulation was done on VEGA cluster of IIT Madras and HPC cluster of IIT Kanpur.

\end{document}